\documentstyle[aps,prl,twocolumn,psfig]{revtex}

\begin{document}
\draft                       
\title{Random walks on fractals and stretched exponential relaxation}

\author{Philippe Jund$^1$, R\'emi Jullien$^1$ and Ian Campbell$^{1,2}$}

\address{
         $^1$Laboratoire des Verres,  Universit\'e Montpellier 2,
         place E. Bataillon, 34095 Montpellier, France.
         $^2$Laboratoire de Physique des Solides, Universit\'e Paris-Sud,
         Centre d'Orsay, 91405 Orsay, France.}

\maketitle

\begin{abstract}

Stretched exponential relaxation ($\exp{-(t/\tau)}^{\beta_K}$) is 
observed in a large variety of systems but has not been explained so far. 
Studying random walks on percolation clusters in curved spaces whose 
dimensions range from $2$ to $7$, we show that the relaxation is accurately a 
stretched exponential and is directly connected to the fractal nature of these 
clusters. Thus we find that in each dimension the decay exponent $\beta_K$ is 
related to well-known exponents of the percolation theory in the corresponding 
flat space. We suggest that the stretched exponential 
behavior observed in many complex systems (polymers, colloids, glasses...) 
is due to the fractal character of their configuration space.

\end{abstract}


The stretched exponential decay function, $q(t) =
\exp{-(t/\tau)}^{\beta_K}$, was initially
proposed empirically in 1854 by R. Kohlrausch\cite{kohlrausch} to
parametrise the discharge of Leyden jars, and was
rediscovered in 1970 by Williams and Watts\cite{WW}. Since then the
phenomenological "KWW" expression has been shown to give an excellent
representation of experimental and numerical relaxation data in a huge
variety of complex systems including polymers, colloids, glasses, 
spin glasses, and many more. This behavior has attracted considerable
curiosity;
it has been discussed principally in terms of models where individual
elements relax
independently with an appropriate wide distribution of relaxation times
(see for instance\cite{palmer,sher,phillips,bunde}).
However, despite the ubiquity of the KWW expression, in the view of many
scientists
its status remains that of a convenient but mysterious phenomenological
approximation having no fundamental physical justification.\\
Here we demonstrate that on the contrary KWW is in fact a {\it bona fide}
and respectable relaxation function; exactly this form of relaxation
appears naturally when we consider random walks on fractal structures in
closed spaces of general dimensions. We suggest that the physical
significance of the relaxation behavior in numerous complex systems should be
reconsidered in the light of this result.\\
For random walks on fractal clusters in Euclidean (flat) spaces, 
it is well known
that if $r(t)$ is the distance of the walker from the starting
point after time $t$, then 
$<r^2(t)>\ \propto t^{2/2+\theta} = t^{\beta_{RW}}$\cite{alexander,gefen}.
Here $\beta_{RW}= \tilde d / D$
where $\tilde d$ and $D$ are the spectral and fractal dimensions of the
cluster respectively. For a critical percolation fractal,  $\theta = (\mu -
\beta)/\nu$
where $\mu$, $\beta$ and $\nu$ are universal percolation critical exponents
\cite{alexander,gefen} whose numerical values are known quite accurately in
all
dimensions \cite{adler1,adler2,stauffer,ballesteros1,ballesteros2,lorenz}.\\
For random walks on the surface of a sphere, which is a closed surface, the 
local behavior $<r^2(t)>\ \propto t$ can be shown to lead exactly to an 
exponential decay  of the autocorrelation $<cos(\theta(t))>\propto 
exp(-t/\tau)$ where $\theta(t)$ is the angle between the initial $t=0$ 
position vector of the walker and the position vector at time $t$. With 
an appropriate definition of $<cos(\theta(t))>$ this result holds for 
hyperspheres in any dimension.
For random walks on a percolation fractal inscribed on a {\it closed}
space with the topology of a hypersphere, it was conjectured some years ago
that the end-to-end autocorrelation function should decay as a KWW stretched
exponential, with the Kohlrausch exponent $\beta_K$ equal to the flat space
percolation fractal random walk exponent, i.e. $\beta_K  =
\beta_{RW}$  in any space dimension \cite{campbell1,campbell2}. 
This can be understood simply in terms of a "fractal" time $t^\beta_{RW}$ 
replacing $t$ in both the local $<r^2(t)>$ and the exponential 
$<cos(\theta(t))>$ expressions above.
This conjecture has been
extensively tested numerically but only in the extreme case of the very
high dimensional hypercube \cite{campbell2,lemke,dealmeida} (the hypercube has
the same closed space topology as a hypersphere). In the present
work we have studied numerically the general case of random walks on
percolation clusters on hyperspherical surfaces for the range of different
embedding dimensions $d$ running from $3$ to $8$.\\
The simplest case to visualize is the surface of a sphere in dimension
$d=3$. The surface is decorated with small disks 
whose centers are distributed
at random. Several  clusters, made of overlapping disks, can be determined.
The largest cluster contains more and more disks as
the total number of disks is increased.   
Just as there appears a percolation cluster (containing a non-infinitesimal
fraction of the total number of disks) 
for the equivalent system in the two dimensional flat space above a critical
value of disk concentration, so the largest cluster ``percolates'' on the
surface of the sphere when the number of disks is sufficiently large.
As an illustration, in figure 1, we have represented the two dimensional
projection of the percolating cluster in the case $\delta/R = 0.01$ (where
$\delta$ is the disk diameter and $R$ the sphere radius).
Here $N_p=179200$ disks have been
disposed at random; only the largest cluster, containing 38130 disks, 
is represented. Note that this cluster spans almost a hemisphere,
a situation intermediate between a well localized cluster ($N<<N_p$)
and a cluster spanning uniformly the whole surface of the sphere ($N>>N_p$).\\
The position of a given disk center can be defined by the $d=3$ coordinates
$x_i$, where the origin is taken as the center of the sphere.
Imagine now a walker jumping at random from one disk center
to the center of any disk overlapping it. While the values of the
coordinates averaged over many walks $<x_i(t)>$ stay finite for $N<<N_p$
(the walker is localized) they decay exponentially to zero for 
$N>>N_p$, finally losing memory of their initial values. 
Of course, in the limit  $N>>N_p$, $<x_i(t)>$ goes to zero since the
random walker starts to investigate the largest cluster (which fills
uniformly the curved space) entirely.
Our calculations show
that right at the percolation where the largest cluster is fractal, 
the decay is critical, and takes up precisely the 
stretched exponential form, with a $\beta_K$ exponent equal to $\beta_{RW}$,
already known for the random-walk on a percolating cluster in a
two-dimensional flat space.
Our calculation is the generalization of this picture over a wide range of 
dimensions, in particular for dimensions larger than $n=6$, where it is
known that $\beta_{RW}$ reaches its mean-field value $1/3$.\\
Consider the surface of a $d$-dimensional (hyper)sphere of unit radius 
which can be defined, using Euclidean coordinates, by
$ \sum_{i=0}^{d} x_i^2 = 1$.
This (hyper)surface is a closed and curved $n$-dimensional 
($n=d-1$) space, $S_n$, on which one can define a
geodesic distance between two points (1) and (2) by $\theta =  \cos^{-1} q$
where $q$ is the scalar product of the end positions, i.e.
$q = {\bf r}_1.{\bf r}_2 = \sum_{i=0}^{d} x_i^{(1)} x_i^{(2)}$.
Of course, this distance becomes asymptotically equivalent to the Euclidean
distance in the limit of distances 
infinitesimally small compared to the radius of the (hyper)sphere (which is
here set to  unity). On this $n$-dimensional (hyper)surface,
 identical $n$-dimensional small (hyper)disks  (called disks in the 
following) of diameter $\delta <<1$ are disposed sequentially, the 
successive disk centers being chosen at random uniformly on $S_n$. 
To determine the cluster structure, as soon as a new disk is added,
a search for connections with previous disks is performed by checking if
their  center-to-center geodesic distance is smaller than $\delta$.
At each stage, when $N$ disks have been disposed on $S_n$, we can define 
and label the different clusters made up of connected disks. In particular 
one can determine the largest cluster
containing the greatest number $N_m$ of disks.          
Also one can characterize the overall filling of the space by a
dimensionless parameter $\eta$,
ratio between the sum of the area of the individual disks and the total
area of $S_n$ \cite{technique}. 
Note that this filling parameter is larger than the true volume fraction $c$,
because of multiple counting due to overlaps, and
therefore $\eta$ can eventually exceed
unity. More precisely it has been shown 
that $c=1-\exp{(-\eta)}$\cite{quintinilla,rintoul}.
Percolation occurs for $\eta$ larger than a threshold value $\eta_p$ 
above which $P=N_m/N$, which measures
the probability for  a given disk  to belong to the largest cluster,
remains non-zero in the thermodynamic limit $\delta \rightarrow 0$.
In practice the curve $P(\eta)$ exhibits a sigmo\"\i dal shape which becomes
sharper and sharper at $\eta_p$ as $\delta \rightarrow 0$.\\
Once  the largest cluster has been identified, we choose one of its
constitutive disk centers ${\bf r}(0)$ at random
as a starting point. We perform a random walk on the cluster by performing
successive jumps, first from  ${\bf r}(0)$ to the center  ${\bf r}(1)$ of 
any other disk connected to it (chosen at random over all its 
overlapping disks),
then iterating from ${\bf r}(1)$ to ${\bf r}(2)$, etc..\cite{jump}
After t steps, the ``correlation function'' ${\bf
r}(0).{\bf r}(t)$ is calculated.
This is the scalar product of the two end positions, which is no more than
the cosine of the geodesic end-to-end distance $\theta$ measured on $S_n$.
In practice we calculate the quantity $q(t) = <\cos\theta>$ which has been 
averaged, for a given number of steps $t$, over $N_r$ independent realizations
of the largest cluster as well as over $N_s$ independent
choices of the starting point on each cluster. Given dimension $n$
and size $\delta$, the behavior of $q(t)$ has been analyzed
for different filling values $\eta$. 
\\An example with $n=3$ and $\delta=0.02$ is
shown in figure 2, where $-\ln|q(t)|$
has been plotted as a function of $t$ in a log-log plot, after taking an
average over independent walks starting from $N_s=10500$ different disks 
on each of the $N_r=20$ independently generated largest clusters 
(a typical value of the number of disks in the percolating cluster
is $N_m\simeq 700000$).  
\\If the relaxation function $q(t)$ is strictly a stretched exponential, this
type of plot produces a straight line of slope $\beta_K$.  
For the critical value of $\eta$
(here $\eta_p\simeq 0.3452$), one observes a clear straight line behavior 
in the numerical data over (at least) six decades in $t$ (the walk 
for $\eta_p$ has been extended to illustrate this point).
In the inset, we show how we estimate $\eta_p$ and $\beta_K$. The effective
slope of the log-log plot
has been determined by a least-square fit within an interval of one-tenth
of the total range in $\ln t$
and plotted as a function of $\ln t$. The percolation threshold $\eta_p$ is
estimated as the $\eta$-value giving the widest plateau at large times, 
and the $\beta_K$ exponent is taken as this plateau value. 
This procedure has been repeated for different values of $\delta$ and for
dimensions $n$ ranging from 2 to 7 (for $n$=1 it is well known
that the percolation transition does not occur \cite{stauffer}). The lowest 
attainable $\delta$ values are mainly determined  by the limited memory 
of our computers.            
In practice the values of $N_s$ have been chosen to be of the order of two 
percent of $N_m$, and $N_r$ has been
chosen so as to obtain runs of the order of a day on regular high speed PC
computers. $N_s$ varies from few thousands
to a few units when decreasing $\delta$ from about 0.3 to its  lowest
attainable $\delta$ value.
This protocol leads to error bars of the order of 0.01 for the 
exponent estimates. Of course the ``true'' values of the
percolation thresholds and exponents are  obtained by an extrapolation
to the "thermodynamic" limit $\delta\rightarrow 0$.\\
The numerical results for $\eta_p$ in $n=2$ and $3$ are in excellent
agreement with accurate estimates from flat space calculations
\cite{quintinilla,rintoul}, and $\eta_p$ drops quasi-exponentially with $n$
for higher dimensions. The data for $\beta_K$ are summarized in figure 3,
where $\beta_K$ has  been plotted as a function of $\delta$.
On the $\delta=0$ axis of figure 3, we have indicated the
best  estimates of the flat space $\beta_{RW}$ with their associated error
bars, calculated from recent
$\mu$, $\beta$ and $\nu$ values available in the literature
\cite{adler1,adler2,stauffer,ballesteros1,ballesteros2,lorenz}.
We note that for all $n\ge 6$, $\beta_{RW} = 1/3$ exactly, since 
$n=6$ is the upper critical dimension for percolation\cite{stauffer}. It is 
quite remarkable that for each dimension $n$, a simple straight line fit 
of our data goes through the corresponding $\beta_{RW}$ value to within 
the numerical error or, at least, extrapolates to a value very close to it.\\
It should be noticed that the law $<\cos\theta> \propto 
\exp{-(t/\tau)^{\beta_K}}$ not only contains the large time relaxation
behavior but also the short time behavior as, after expanding both sides 
for small $t$ and $\theta$,
it becomes $<\theta^2> \propto (t/\tau)^{\beta_K}$. Since for short times
the random walker stays on a $n$-dimensional surface tangent to the 
(hyper)sphere, one recovers
the law $<r^2> \propto t^{\beta_{RW}}$ in dimension $n$. This could
explain why the stretched exponential
behavior extends over so large a region of time (see figure 2). In
practice, as in flat space, at very short times there are          
corrections to scaling  due to the discrete character of the walk.\\   
These data can be taken as a clear numerical demonstration
that  random walks on a fractal cluster inscribed on a hypersphere in any 
dimension lead necessarily to a stretched exponential decay of the 
correlation function, with a Kohlrausch exponent $\beta_K$ equal to the ratio 
${\tilde d} / D$ of the spectral and fractal dimensions of the cluster. 
Thus the stretched exponential relaxation on a fractal in a closed space 
appears as the precise analogue of the sub-linear diffusion on a fractal in a 
flat space.\\
Why is this relevant to relaxation in complex systems ? A complex system is
made up of many individual elements (atoms, molecules, spins,...etc),
all in interaction with each other. The total space of all possible
configurations of the whole system is a huge closed space 
having a very high dimension, of the order of the number of elements. 
These spaces are so astronomically large that an explicit evaluation of their 
properties, configuration by configuration, is almost impossible except 
for tiny systems. Each configuration has an energy associated to it. At 
finite temperature $T$ only a restricted subset
of configurations are of low enough energy to be thermodynamically
accessible by the system. In equilibrium at $T$ above any ordering
temperature, the system is permanently
exploring all this subset of accessible configurations by successive 
movements or reorientations of local elements of the total system. 
By definition this equilibrium relaxation can be mapped 
onto a random walk of the point representing the instantaneous
configuration of the total system within the space of accessible
configurations. 
Thus the configuration space of a complex system can be viewed as 
a ``rough landscape''. In such a scenario, the portion of configuration 
space available to the system consists of only a restricted set of 
tortuous configuration space paths. In this situation,
when real measurements are made, the observed relaxation functions must 
reflect the complex morphology of the available
configuration space, and so will be slow and non-exponential - typically
stretched exponential. 
Now, any relaxation function including the KWW function can be represented as
the sum of an appropriate distribution of 
elementary exponential relaxations. However it is important to note that the 
independent exponentially relaxing elements in a complex system are the {\it
modes} of the whole system not the atoms, molecules or spins which are in 
strong interaction with each other\cite{dealmeida}. 
It is the morphology of the configuration space which determines the 
mode distribution and the form of the relaxation.\\
We have just seen that a closed space fractal structure
necessarily leads to a relaxation process which is exactly of 
stretched exponential type; we suggest that, inversely, when a complex 
system at temperature $T$ is actually observed to relax
with a stretched exponential, it is the signature of a 
fractal morphology of the available configuration space at that temperature. 
In particular, whatever the microscopic details of the interactions 
in a glass former are, it is generally observed that as the glass temperature 
is approached from above, KWW relaxation sets in. The implication is that
phase space takes up a fractal morphology as a consequence of the 
intrinsic complexity associated
 with glassiness. The image of the glass transition which follows is 
that of a percolation transition in phase space\cite{campbell1,dealmeida}
with the temperature being analog to the filling parameter.
As would be expected from the argument given above, the Kohlrausch 
exponent is observed to tend to a limiting value of $1/3$ as the glass 
transition is approached in a number of systems,
\cite{ogielski,bartsch,angelani,alegria} for instance.\\
In summary, we have demonstrated that random walks 
on fractals in closed spaces give stretched exponential relaxation, 
and we suggest that stretched exponential relaxation is ubiquitous 
in nature because configuration spaces are fractal in many complex systems.

%
\hrule
\begin{figure}
\psfig{file=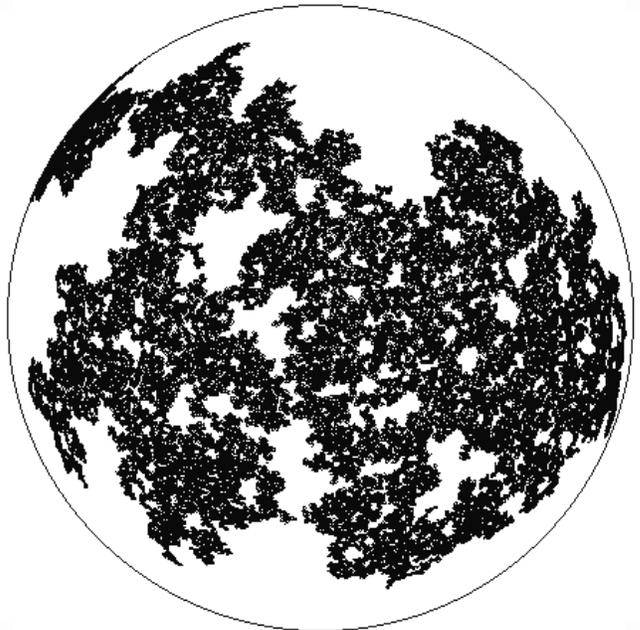,width=8.5cm,height=8.5cm}
\caption{ Two-dimensional projection of a typical percolating cluster of disks
of diameter $\delta=0.01$  on the surface of a sphere of unit radius.
Here $N_p=179200$ disks have been
disposed at random; only the largest cluster, containing $N_m=38130$ disks,
is represented.}
\label{Fig.1}
\end{figure}
\vskip 1cm
\begin{figure}
\psfig{file=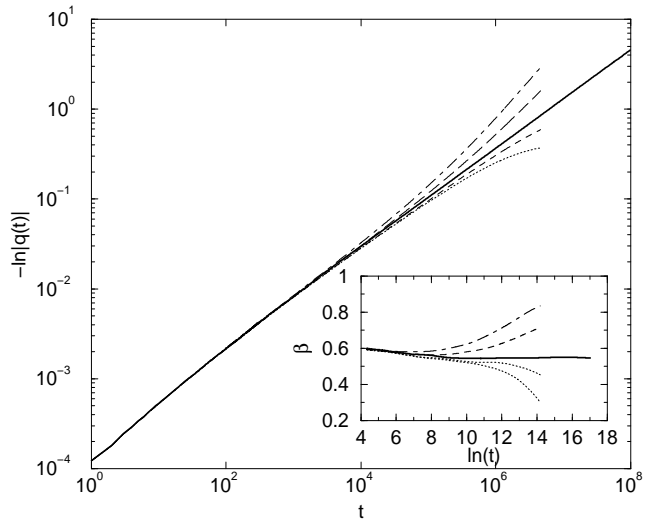,width=8.5cm,height=7cm}
\caption{Plot of $-\ln|q(t)|$ versus t (averaged over $N_r=20$ and 
$N_s=10500$) for $n=3$, $\delta=0.02$ and for different $\eta$ values: 
from bottom to top
$\eta=$ 0.335, 0.34, 03452, 0.35, 0.355 (the statistical error bars do not
exceed the thickness of the lines). In inset the local value of 
the exponent $\beta$ is plotted as a function of $\ln t$ for the different $\eta$ values in
order to estimate $\eta_p$ and $\beta_K$.}
\label{Fig. 2}
\end{figure}

\begin{figure}
\psfig{file=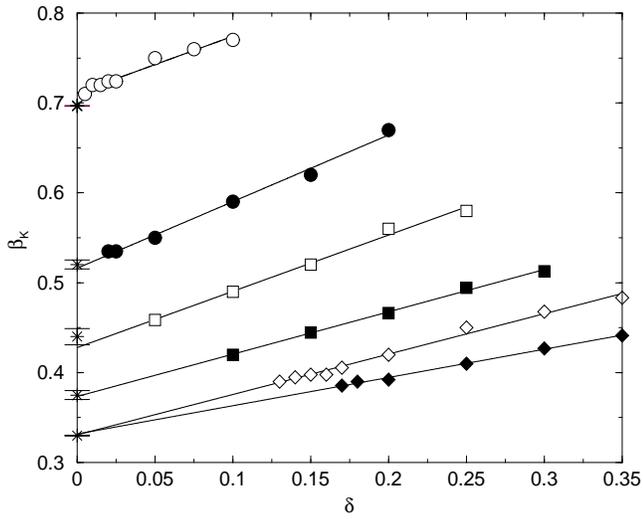,width=8.5cm,height=7cm}
\caption{Variation of the exponent $\beta_K$ as a function of the (hyper)disk 
diameter $\delta$ for
dimension $n=2,3,4,5,6$ and $7$ (from top to bottom). 
The solid lines represent least square linear
fits of the data points. The best estimates of the flat space exponent 
$\beta_{RW}$ with their associated error bars [9-14] are represented by 
the stars on the y axis.}
\label{Fig. 3}
\end{figure}


\begin{references}
\bibitem{jump} Note that the jump lengths are uniformly distributed between 0 
and $\delta$. It is a well known result of the percolation theory that the 
precise form of the length distribution (as soon as it is bounded)
 is irrelevant for the determination of the critical exponents.
\bibitem{technique} $\eta$ is defined  by
$\eta = N {\omega_n (\delta/2)\over \omega_n(\pi)}$
where $\omega_n(\theta)$ is the area of  a (hyper)disk of radius $\theta$.
Using standard algebra, one gets
$\eta  = N {I_{n-1}(\delta/2)\over I_{n-1}(\pi)}$ with
$I_n(\theta) = \int_0^\theta \sin^n\phi d\phi$
\end{references}
\end{document}